\documentclass[twocolumn,showpacs,amsmath,amssymb,prb]{revtex4}
\usepackage{dcolumn}% Align table columns on decimal point
\usepackage{epsfig}
\usepackage[latin1]{inputenc} 
\usepackage{bm}% bold math
\begin{document}

%\documentstyle[eqsecnum,multicol,epsfig,aps,prb,array]{revtex}
%\newcommand{\bleq}{\ifpreprintsty
%		   \else
%		   \end{multicols}\widetext \vspace*{-3.5ex}{\tiny
%		   
%		\noindent\begin{tabular}[t]{c|}
%		   \parbox{0.493\hsize}{~} \\ \hline \end{tabular}}
%				      \fi}
%\newcommand{\eleq}{\ifpreprintsty
%		   \else
%		   {\tiny\hspace*{\fill}\begin{tabular}[t]{|c}\hline
%		    \parbox{0.49\hsize}{~} \\
%		    \end{tabular}}\vspace*{-2.5ex}\begin{multicols}{2}
%		    \narrowtext
%		    \fi}
%\newcommand{\bcols}{\ifpreprintsty\else\begin{multicols}{2} 
%	\narrowtext\fi}
%\newcommand{\ecols}{\ifpreprintsty\else\end{multicols}\fi}
%
%
%\draft
%\widetext
%\begin{document}
\title{Fast-ion conduction and flexibility and rigidity of solid electrolyte glasses} 
\author{M. Micoulaut$^1$, M. Malki$^{2,3}$ , D.I. Novita$^4$, P. Boolchand$^4$}
\affiliation{$^1$ Laboratoire de Physique Théorique de la Matière Condensée,
Université Pierre et Marie Curie,  Boite 121, 4, Place Jussieu, 75252
Paris Cedex 05, France\\
$^2$ CEMHTI,CNRS, UPR 3079 1D Avenue de la recherche scientifique, 45071 Orléans Cedex 02, France\\
$^3$ Université d'Orléans (Polytech' Orléans), BP 6749, 45072 Orléans Cedex 02, France\\
$^4$ Department of Electrical and Computer Engineering, University of Cincinnati, Cincinnati, OH 45221-0030, USA}

\date{\today}
%\maketitle
\begin{abstract}
Electrical conductivity of dry, slow cooled (AgPO$_3$)$_{1-x}$(AgI)$_x$ glasses  is examined as a function of temperature , frequency and glass composition. From these data compositional trends in activation energy for conductivity E$_A$(x), Coulomb energy E$_c$(x) for Ag$^+$ ion creation, Kohlrausch stretched exponent $\beta$(x), low frequency ($\varepsilon_s$(x)) and high-frequency ($\varepsilon_\infty$(x)) permittivity are deduced. All parameters except E$_c$(x) display two compositional thresholds, one near the stress transition, x = x$_c$(1)= 9\%,  and the other near the rigidity transition,  x = x$_c$(2)= 38\% of the alloyed glass network. These elastic phase transitions were identified in modulated- DSC, IR reflectance and Raman scattering experiments earlier.  A self-organized ion hopping model (SIHM) of a parent electrolyte system is developed that self-consistently incorporates mechanical constraints due to chemical bonding with carrier concentrations and mobility. The model predicts the observed compositional variation of $\sigma$(x), including the observation of a step-like jump when glasses enter the Intermediate Phase at x$>$x$_c$(1),  and an exponential  increase  when glasses become flexible  at x$>$x$_c$(2). Since E$_c$ is found to be small compared to network strain energy (E$_s$), we conclude that free carrier concentrations are close to nominal AgI concentrations, and that fast-ion conduction is driven largely by changes in carrier mobility induced by an elastic softening of network structure.  Variation of the stretched exponent $\beta$(x) is square- well like with walls localized near x$_c$(1) and  x$_c$(2) that essentially coincide with those of the Intermediate Phase (IP) (x$_c$(1)$<$x$<$x$_c$(2)),  and suggest filamentary (quasi 1D) conduction in the IP, and conduction with a dimensionality greater than 1 outside the IP. 
\end{abstract}
\maketitle
%\bcols 

\section{Introduction}

Fast-ion conduction in solid electrolyte glasses has been investigated
 \cite{1}-\cite{6} for over two decades and our basic understanding continues to be challenged with new findings \cite{7}-\cite{9} in the field. Resolution of these basic challenges \cite{10} including the role of molecular structure and impurities \cite{11} on fast-ion conduction is likely to have a direct impact on applications of these materials in solid state batteries, sensors, and non-volatile memories.
\par
Electrical conductivity $\sigma$ can be generally be written as 
\begin{eqnarray}
\label{eq1}
\sigma=Ze\mu n_L
\end{eqnarray}                                                              
where the terms  $Ze$, $\mu$, $n_L$ represent respectively the charge of the conducting ion, its  mobility and  concentration. Beyond this point of discussion, it is clear that alloying ionic species in a base glass will lead to an increase in carrier concentration in a host network, but it will also substantially modify the mechanical behaviour by either depolymerising or softening the base structure. Solid electrolyte additives such as AgI, Ag$_2$S and Ag$_2$Se in base chalcogenide glasses are known to possess a low connectivity and their glass forming tendency derives from that feature of structure \cite{12}-\cite{13}. Recently Ingram et al. have proposed \cite{10} that the activation energy for conduction (E$_A$) could be displayed under the form of a master plot with respect to activated volume (V$_A$) as 
\begin{eqnarray}
\label{eq2}
E_A=MV_A
\end{eqnarray}
where $M$ represents a localized modulus of elasticity (strongly influenced by intermolecular forces) with values ranging between 1 and 8 GPa depending on the nature of chemical bonding  involved.  One may, therefore, expect that ionic conduction can be related to an elastic softening of a glass network with one functionality (elasticity) affecting the other (conductivity). Can this view be made more quantitative?  
\par
Conductivity thresholds in fast ionic conductors have been reported \cite{14} although the relationship with rigidity transitions was not made. Micro-segregation effects have also been reported \cite{15} in various modified oxide glasses including silicates and supported \cite{16} by emerging length scales in MD simulated partial pair distribution functions. These findings are popularized by the channel picture for alkali ion-conduction in sodium silicates due to Greaves and Ngai \cite{14}. In this glass system, the soda content where a micro-segregation onsets is nearly the same as the stress elastic phase transition (x = 18\%) determined from calorimetric measurements \cite{17}. A natural question that follows then is, are there links between the elastic behaviour of glasses and micro-segregation effects? Is this a generic behaviour or is it typical of alkali-modified silicates? 
\par
(AgPO$_3$)$_{1-x}$(AgI)$_x$ glasses  have been widely studied in the past by several groups \cite{18}-\cite{25}. A perusal of this large body of work reveals broad compositional trends in thermal, optical and electrical properties of these materials, but not without significant variations between various groups. In some case the conductivity increases monotonically with AgI content displaying either no threshold \cite{26} or one threshold near x = 30\% \cite{27}-\cite{28}. In more recent reports \cite{8}$^,$ \cite{11}$^,$ \cite{29}$^,$\cite{30} we have emphasized the need to keep samples dry during synthesis and handling, and to slow cool them from T$_g$ to assure measuring what we believe are the intrinsic physical properties of these materials. The thermal, optical, molar volumes and some electrical properties were reported \cite{30}, and these are found to be quite different from those reported earlier in the literature \cite{20}$^,$ \cite{26}$^,$ \cite{28} in that many properties including conductivity display two striking compositional thresholds.  The data from these more recent investigations \cite{8}$^,$ \cite{30} examined as a function of glass composition are found to display striking correlations between various observables. For example, these data have permitted connecting aspects of local structure deduced from Raman and IR vibrational spectroscopy to their thermal as well as electrical transport properties. The elastic nature of backbones derived from Lagrangian constraints, have permitted treating short-range intermolecular forces in an effective way using rigidity theory.
\par
\begin{figure}
\begin{center}
\epsfig{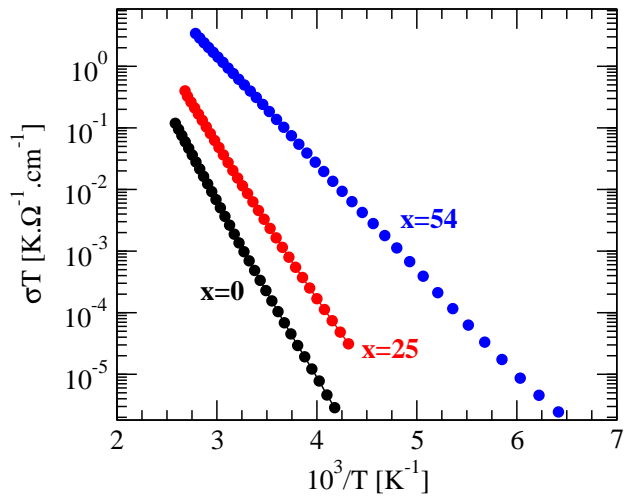}
\end{center}
\caption{(Color online) Plot of  electrical conductivity ($\sigma$) multiplied by T as a function of 1000/T at the three indicated  AgI content x in  (AgPO$_3$)$_{1-x}$-(AgI)$_x$ glasses displaying an Arrhenius variation.}
\end{figure}
 In the present contribution we present temperature dependent and frequency dependent conductivity data on these glasses as a function of AgI content x. Here we address an issue of major importance in the field of solid electrolytes, namely, the relationship of the three defined elastic phases: flexible, intermediate, stressed rigid to the dielectric permittivity derived from the frequency-dependent electrical modulus. We find that (a) the dielectric permittivity is a maximum in the intermediate phase (IP), (b) the Kohlrausch stretched exponent $\beta$ displays a minimum there and (c); we also infer there are  dimensional changes in conduction pathways   with changing glass composition, and (d)  that Coulomb barrier , E$_c$, determined from the high-frequency permittivity  makes only a small contribution to the Arrhenius activation energy for dc conduction, E$_A$,  compared to network strain energy, E$_s$. These findings suggest that free carrier concentrations are nearly equal to one, and that they closely follow the AgI concentrations. The observed compositional trends in dc conductivity, $\sigma$(x) appear to derive from elastic network softening which controls carrier mobility. The observed precipitous increase of conductivity in the flexible phase may well be a generic feature of solid electrolyte glasses.
 
\section{Experimental}

\subsection{Synthesis of dry AgI-AgPO$_3$ glasses}
Bulk (AgPO$_3$)$_{1-x}$(AgI)$_x$ glasses were synthesized using 99.9\% Ag$_3$PO$_4$ (Alpha Aesar Inc.),  99.5\% P$_2$O$_5$ ( Fischer Scientific Inc.), and 99.99\% AgI (Alpha Aesar Inc.), and separately AgNO$_3$ and NH$_4$H$_2$(PO$_3$)$_4$ with AgI as the starting materials. Details of the synthesis are discussed elsewhere \cite{11}. In our experiments, the starting materials were handled in a dry nitrogen gas purged glove box (Vacuum Atmospheres model HE-493/MO-5, relative humidity $\ll $ 0.20\%). The starting materials were weighed in the desired proportion and thoroughly mixed in an alumina crucible with all handling performed in the glove box.  Mixtures were then transferred to a box furnace held at 125$^o$C in a chemical hood purged by laboratory air, and heated at 100$^o$C/hr to 700$^o$C. Melts were equilibrated overnight and then quenched over steel plates. Calorimetric experiments also revealed \cite{11} that stress frozen upon melt quenching can be relieved by cycling samples through the glass transition temperature (T$_g$). Conductivity measurements were undertaken only on samples cycled through T$_g$ and slow cooled at 3$^o$C/min. to room temperature.

\subsection{AC conductivity measurements as a function of temperature and frequency}

   Glass sample disks about 10mm in diameter and 2 mm thick were synthesized by pouring melts into special troughs. Pellets were then thermally relaxed by cycling through T$_g$. Platelets then were polished, and the Pt electrodes deposited. The complex impedance Z$^*$($\omega$)=Z'($\omega$)+iZ''($\omega$) of the specimen was measured by a Solartron SI 1260 impedance analyzer over a frequency range of 1 Hz-1 MHz (19 points per measurements). The temperature was raised from 150 K to T$_g$ at a rate of 1$^o$C/min.  All data were acquired automatically at 2 min intervals. The complex conductivity $\sigma^*$($\omega$)=$\sigma$'($\omega$)+i$\sigma$''($\omega$) is deduced from the complex impedance Z$^*$, the sample thickness t and the area S of the surface covered by platinum using the formula : $\sigma^*$($\omega$)=(t/S)(1/Z$^*$).
    
\section{Experimental results}

\subsection{Temperature dependence of conductivity}

The temperature dependence of electrical conductivity for ionic conduction in glasses usually displays an Arrhenius dependence 
\begin{eqnarray}
\label{eq1}
\sigma=\sigma_0\exp[-E_A/k_BT]
\end{eqnarray}
We have obtained the activation energy E$_A$(x), by performing temperature dependent measurements in the 40$^o$C$<$T$<$T$_g$ range. An illustrative example of some of these results, at three glass compositions, is reproduced in Fig 1.  From the temperature dependence, we have obtained the preexponential factor $\sigma_0$ and the activation energy E$_A$ as a function of AgI content, and these data are reproduced in Fig.2. One finds that E$_A$(x) steadily decreases as x increases, particularly for x $>$ 40\% while $\sigma_0$ drops rapidly with x at low x ($<$10\%) and then levels off in the intermediate x region.
\par
Interestingly, we find $\sigma$(x) displays three regimes of conduction whatever the temperature (Fig. 3a). At low x ($<$ 9\%) $\sigma$(x) is found to be nearly independent of x. The conductivity then builds up at higher x ($>$ 10\%) with a distinct step near 9\%, and at 
x $>$ 37.8\% a power-law variation of $\sigma$(x) sets in. The location of the two electrical conductivity thresholds at x = 9\% and x = 38\% coincide with steps in the non-reversing enthalpy at T$_g$ that define the boundary of the Intermediate Phase (IP) (Figure 3b).  These data will be discussed in section IV.
\begin{figure}
\begin{center}
\epsfig{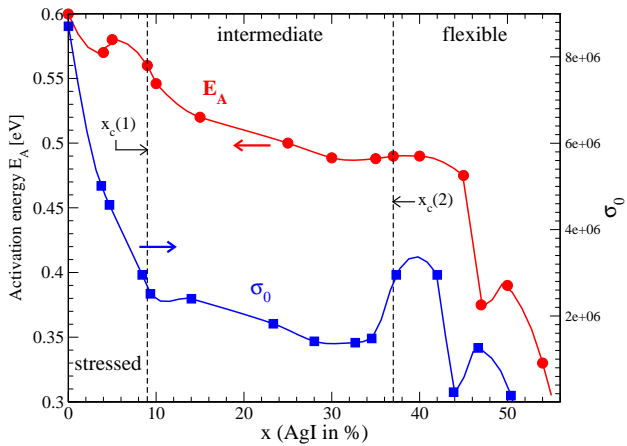}
\end{center}
\caption{(Color online) Variations in activation energy E$_A$(x) (blue) and preexponential factor $\sigma_0$(x) (red) as a function of AgI content x in percent in (AgPO$_3$)$_{1-x}$-(AgI)$_x$  glasses.}
\end{figure}

\begin{figure}
\begin{center}
\epsfig{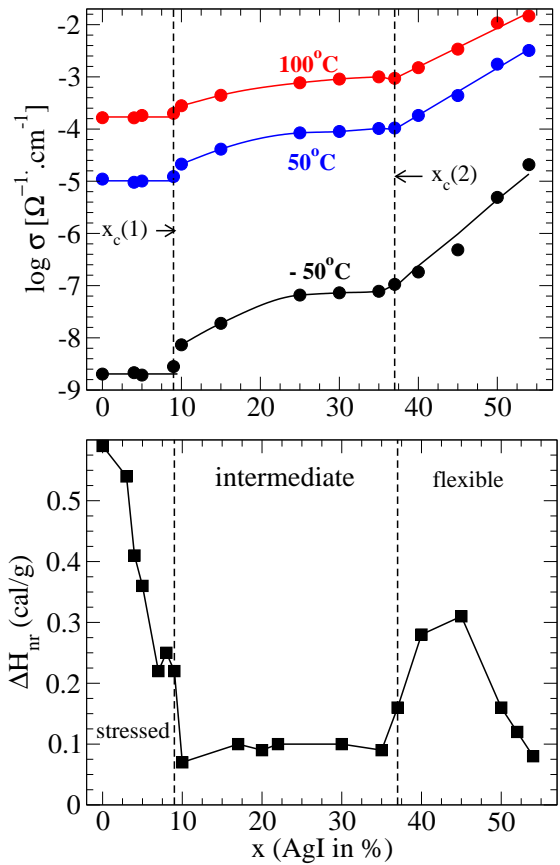}
\end{center}
\caption{(Color online) a: Variations in DC conductivity $\sigma$(x) at three temperatures, -50$^o$C, 50$^o$C and 100$^o$C, and b: Non-reversing heat flow $\Delta$H$_{nr}$ at T$_g$  in  (AgPO$_3$)$_{1-x}$-(AgI)$_x$ glasses as a function of AgI content in mole \%.}
\end{figure}

\subsection{Frequency dependence of AC conductivity}

The frequency dependence of AC conductivity, $\sigma$($\omega$) at three representative glass compositions is summarized in  Figure 4. These data show a saturation of $\sigma$($\omega$) at low frequencies  ($\omega$/2$\pi$ $<$ 100 Hz), and an increase of about 2 orders of magnitude in the 10$^3$ $<$ $\omega$ $<$ 10$^6$ range.  The low frequency saturation value of $\sigma$ is found to increase steadily as the AgI content of glasses x increases from 10\% to 40\%, a behaviour that is consistent with data in Fig. 3. The present results on the frequency dependence $\sigma$($\omega$)  are  similar to earlier reports of Sidebottom \cite{20} and separately by Stanguenec and Elliott \cite{21}. For a more quantitative analysis of these $\sigma$($\omega$) data, we  have deduced the frequency dependence of the dielectric permittivity, $\varepsilon_\infty$($\omega$) from the measured AC conductivity.

\begin{figure}
\begin{center}
\epsfig{figure=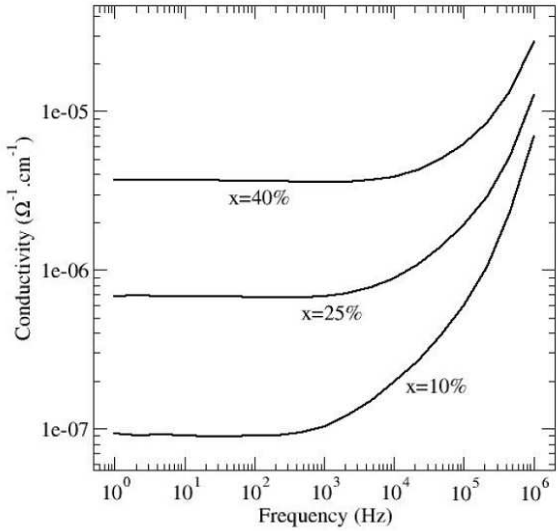,width=0.9\linewidth}
\end{center}
\caption{Frequency dependence of electrical conductivity in  (AgPO$_3$)$_{1-x}$-(AgI)$_x$ glasses at three indicated  AgI concentrations x. Measurements were taken at 250K. }
\end{figure}

\subsection{Electrical modulus analysis}

The behaviour of the complex conductivity  $\sigma^*$($\omega$)=$\sigma$'($\omega$)+i$\sigma$''($\omega$) can be analyzed using the complex permittivity  $\varepsilon^*$($\omega$)=$\varepsilon$'($\omega$)+i$\varepsilon$''($\omega$) or the complex electrical modulus M$^*$($\omega$)=M'($\omega$)+iM''($\omega$) which are related to the complex conductivity via:
\begin{eqnarray}
\label{eq4}
\varepsilon^*(\omega)={\frac {\sigma^*(\omega)}{i\omega\varepsilon_0}}
\end{eqnarray}
and
\begin{eqnarray}
\label{eq4}
M^*(\omega)={\frac {i\omega\varepsilon_0}{\sigma^*(\omega)}}={\frac {1}{\varepsilon^*(\omega)}}
\end{eqnarray}
Here $\varepsilon_0$ designates the permittivity of free space of $8.85\times10^{-12}$ Farad/meter. From the measured value of $\sigma$'', using equation (4) we could deduce the real part of the permittivity $\varepsilon$'. These data are presented on a log-log plot for a glass at a composition x = 25\% in Figure 5.  
\begin{figure}
\begin{center}
\epsfig{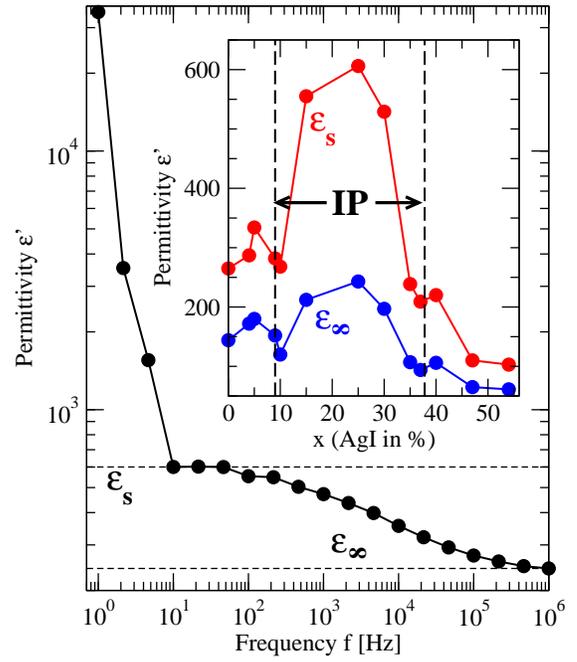}
\end{center}
\caption{(Color online) Log-log plots of the real part $\varepsilon$' of permittivity vs frequency for   (AgPO$_3$)$_{1-x}$-(AgI)$_x$  glasses at x = 25\% measured at 250 K. The broken horizontal lines correspond to the high and low-frequency limits of the permittivity $\varepsilon_\infty$ and $\varepsilon_s$. The inset shows variation of $\varepsilon_\infty$(x) and  $\varepsilon_s$(x) with AgI content. Vertical broken lines define the location of the stress and rigidity transitions at x$_c$(1) = 9\% and x$_c$(2) = 38\% as determined from calorimetry.}
\end{figure}
Permittivity results generally from mobile ions, but the sharp increase in $\varepsilon$' at low frequencies, particularly at f = $\omega$/2$\pi$ $<$ 10 Hz, is due to electrode polarization effects as Ag$^+$ ions are deposited at the anode at very low frequencies \cite{31}. The behaviour is observed for all samples but the magnitude of $\varepsilon_s$ could be unambiguously determined. The saturation value of the permittivity at low frequency ($\varepsilon_s$) and at high frequency ($\varepsilon_\infty$) were thus obtained at all other glass compositions, and these data appear in the  insert of Fig 5. Noteworthy in these trends of $\varepsilon_s$(x) and $\varepsilon_\infty$(x) is the fact  that both show a broad maximum that onsets near x = x$_1$ $\simeq$ 10\% and ends near x = x$_2$ $\simeq$ 35\%, the two  threshold compositions observed in the non-reversing heat flow at T$_g$ (Fig.3b) and electrical conductivity (Fig.3a) that essentially define the IP region \cite{8} of these glasses. At the glass composition x = 20\% , located near  the middle of the IP, these parameters acquire  rather large magnitudes:  $\varepsilon_s$=620 and $\varepsilon_\infty$=250. The magnitude of these permittivity found in the IP of the present electrolyte are truly exceptional- they are high compared to those found in other systems such as the alkali silicates ($\varepsilon_\infty$=11, \cite{14}), and highlight another remarkable property of solid electrolytes bearing Ag salts.
\par
We now follow the approach developed by Ngai and co-workers \cite{31} and use the electrical modulus formalism to obtain stretched exponents $\beta$(x) from our $\sigma$($\omega$) data. Our starting point is to write the electric field E(t) under the constraint of a constant electric displacement as: 
\begin{eqnarray}
E(t)=E(0)\Phi(t)
\end{eqnarray}					
and define E(0) as the  initial electric field and $\Phi$(t) the relaxation function of the system. In the frequency domain, the electrical relaxation is appropriately represented by the electric modulus:
\begin{eqnarray}
M^*(\omega)={\frac {1}{\varepsilon_\infty}}\biggl[1-\int_0^\infty e^{-i\omega t}\biggl(-{\frac {d\Phi}{dt}}\biggr)dt\biggr]
\end{eqnarray}
and the high-frequency ($\varepsilon_\infty$)and low-frequency limits ($\varepsilon_s$) of the real part of the permittivity ($\varepsilon^{'}$) then can be written as :
\begin{eqnarray}
\varepsilon_s=\varepsilon_\infty <\tau^2>/<\tau>^2
\end{eqnarray}
where:
\begin{eqnarray}
<\tau^n>=\int_0^\infty t^{n-1}\Phi(t)dt
\end{eqnarray}
We use the standard relaxation function based on a stretched exponential:
\begin{eqnarray}
\Phi(t)=exp[-(t/ \tau)^\beta]
\end{eqnarray}
with   representing the Kohlrausch-Williams-Watts exponent. The exponent provides a good measure of the correlations of atomic motions in a relaxing network \cite{31}.  And the absence of cooperative behaviour will lead inevitably to a typical Debye relaxation with $\beta$ =1, a behaviour that is expected in dilute limit where concentrations of diffusing cations is low. One finally obtains from equations (8)-(10):
\begin{eqnarray}
\label{beta}
\varepsilon_s=\beta{\frac {\Gamma(2/\beta)]}{[\Gamma(1/\beta)]^2}}\varepsilon_\infty
\end{eqnarray}
where $\Gamma$ represents the Gamma function.
\begin{figure}
\begin{center}
\epsfig{figure=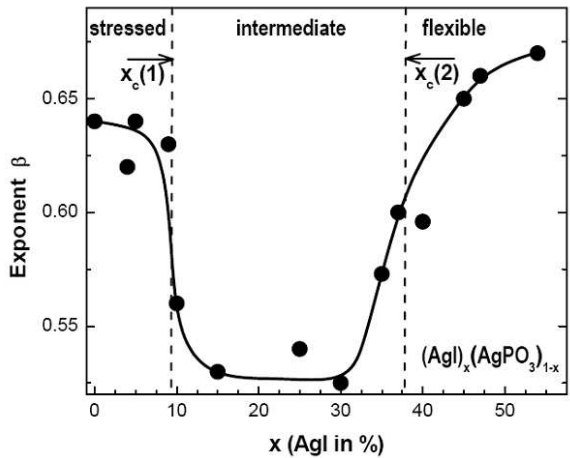,width=0.9\linewidth}
\end{center}
\caption{Variations in the Kohlrausch stretched exponent $\beta$(x)  as a function of the AgI concentration x in (AgPO$_3$)$_{1-x}$-(AgI)$_x$  glasses.}
\end{figure}

The data of Fig. 5 gives the glass composition variation of the permittivity, $\varepsilon_s$(x) and $\varepsilon_\infty$(x). Using these permittivity parameters and the relation between them (equation (11)), we have extracted  compositional variation of the Kohlrausch stretched exponent $\beta$(x)   These data appear in Figure 6. We find $\beta$ $\simeq$ 0.65 at low x (0\%) and at high x (50\%), but at compositions in between, particularly in the IP,  $\beta$ is found to decrease ($\beta$ $\simeq$ 0.53) and to  display a broad square-well like minimum, a trend reminiscent of the reversibility window (Fig.3(b)).  At low carrier concentrations, x $\simeq$ 5\% for example, we are in the dilute limit and carriers can be expected to be  randomly distributed displaying little or no correlations in their motion. Yet our data shows that $\beta$ does not acquire a value of  1, as found in the sodium silicates \cite{14}. This is a curious result and it possibly reflects the fact that in the present system the diffusing cation of interest Ag$^+$ also happen to be part of the base glass network.

\section{Discussion}

\subsection{Three regimes of fast-ion conduction and elastic phases of AgPO$_3$-AgI glasses}

Perhaps the central finding of this work is that fast-ion conduction in the present solid electrolyte glass displays three distinct regimes of behaviour, and that these regimes coincide with the three known elastic phases of the backbone. The result emerges from the compositional trends in the electrical and thermal data presented earlier in Figure 3. The observation of a reversibility window fixes the three elastic phases of the present glasses \cite{8} with glass compositions at x $<$ x$_1$ = 9\% belonging to the stressed-rigid phase, those at x $>$ x$_2$ = 38\% to the flexible phase, while those in between (x$_1$ $<$ x $<$ x$_2$) belonging to the IP. The two thresholds, x$_1$ and x$_2$, we identify respectively with the stress and the rigidity transitions. As glasses become flexible at x $>$ 38\%, $\sigma$(x) increases logarithmically, a result that underscores the crucial role of network flexibility in opening a doorway for Ag$^+$ ions to migrate. In the flexible phase, network backbones can be more easily elastically deformed (less cost in strain energy), and ions can more easily diffuse and contribute to conduction. 
\par
A closer inspection of Fig. 3a also shows that $\sigma$(x) is nearly independent of x in the stressed-rigid glasses ( x $<$ x$_1$), but it varies discontinuously displaying a jump near the stress transition, x$_1$ $\simeq$ 9\%. The jump in $\sigma$(x) decreases as temperature is increased from -50$^o$C to 100$^o$C. At low temperature (-50$^o$C), we note that the observed jump is rather large; $\sigma$ increases by a factor 2.5 between the composition x =10\% and x = 10.5\%. The discontinuity in $\sigma$(x) at the stress transition is reminiscent of a similar behaviour encountered in the variation of the Raman mode frequency ($\nu$) of corner-sharing tetrahedra in binary Ge$_x$S$_{1-x}$ glasses at the stress transition \cite{32}. The frequency jump ($\Delta\nu$) across the threshold also increases as T is lowered.  Since the stress transition is found to be a first order transition, it suggests that the first derivative of the free energy must be temperature dependent.

\subsection{Self-organized ion hopping model (SIHM) of solid electrolytes}

To gain insights into the origin of the observed electrical conductivity thresholds (Fig. 3), we use Size Increasing Cluster Approximation \cite{33}-\cite{34} (SICA) to identify the two elastic thresholds in a parent solid electrolyte glass system. For this purpose we use the alkali oxide modified Group IV oxide, (1-x)SiO$_2$-xM$_2$O, with M=Li,Na,K, and start with  N tetrahedra (e.g. SiO$_{4/2}$ and MSiO$_{5/2}$ (sharing one non-bridging or terminal oxygen) with respective probabilities (1-p) and p=2x/(1-x) at the first step, l = 1, in the agglomeration process. Our goal is to compute the probability of finding specific clusters in the three phases of interest (flexible, intermediate, stressed rigid). We note that a SiO$_{4/2}$ structural unit is stressed rigid (n$_c$=3.67 per atom) while a  MSiO$_{5/2}$ unit is flexible (n$_c$=2.56 per atom). Here n$_c$ represents the count of bond-stretching and bond-bending constraints per atom.  Henceforth, we denote these local structures as St (stressed-rigid) and Fl (flexible) units respectively. Starting from these local units as building blocks, we calculate all possible structural arrangements to obtain agglomerated clusters containing two (step l=2), and then three building blocks (l=3), etc. and their corresponding population probabilities p$_i$. In calculating the probabilities of the agglomerated clusters we fold in their mechanical energies. Details of the method and application appear elsewhere \cite{33}. At each agglomeration step l, we compute the floppy mode count f$^{(l)}$. At  step l=2, the count of floppy modes f$^{(2)}$, 
\begin{eqnarray}
\label{eqnc}
f^{(2)}=3-n_c^{(2)}=3-{\frac {\sum_{j,k=Fl,St} n_{c(jk)}p_{jk}}{\sum_{j,k=Fl,St} N_{jk}p_{jk}}}
\end{eqnarray}
where n$_{c(jk)}$ and N$_{jk}$ are respectively the number of constraints and the number of atoms found in a cluster with probability p$_{jk}$. An IP is obtained if self-organization is achieved \cite{33}-\cite{34}, i.e. if upon decreasing the content of modifier atoms (x), stress-free cyclic (ring) structures form. In these calculations we define a parameter $\eta$ which measures the fraction of atoms in ring structures to those in dendritric ones. Thus, $\eta$=0 corresponds to the case of networks composed of only  dendritic structures (no rings), while $\eta$=1, corresponds to networks with only  ring structures. Our calculations reveal that  IP-widths are directly related to the parameter $\eta$ i.e, larger $\eta$ the greater the IP-width. In practice, one starts with a flexible network containing a high concentration (x) of modifier ions. One then decreases the modifier concentration,  and can expect more St species to form in a cluster that still possesses a finite concentration of floppy modes f$^{(2)}>$0. At a certain composition x=x$_r$, there will be enough St species formed to drive clusters to become rigid when f$^{(2)}$ vanishes, thus defining the rigidity transition. And if we continue to reduce the modifier concentration beyond the rigidity transition, one can expect clusters that are almost stress-free by balancing St and Fl units. Such a selection rule of self-organization will hold up to a certain point in composition x=x$_s$ , the stress transition, i.e., at x $<$ x$_s$ dendritic or  St-rich structures will predominate and stress will percolate across clusters.  In this SICA approach, the IP width is then defined by the composition interval, x$_r$-x$_s$ , which may be compared to the experimentally established interval between the observed stress and rigidity transitions, i.e., x$_c$(2) - x$_c$(1) = 28\% (Fig 3b). In what follows, we fix the value $\eta$  in order to work with a fixed IP-width.
\par
\begin{figure}
\begin{center}
\epsfig{figure=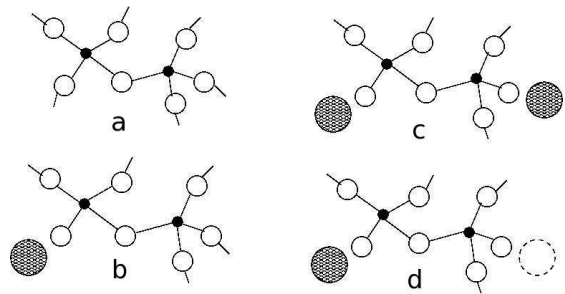,width=0.85\linewidth}
\end{center}
\caption{Structural sites found in the model of (1-x)SiO$_2$-xM$_2$O: a) a stressed rigid St-St connection (n$_c$=3.67) representative of the base glass SiO$_2$, b) an isostatically rigid MSiO$_{5/2}$-SiO$_2$ connection (St-Fl with n$_c$=3.0) containing one M cation, c) a flexible MSiO$_{5/2}$-MSiO$_{5/2}$ connection (Fl-Fl  with n$_c$=2.56) having two M cations. Both b) and c) can give rise to vacant sites such as found on the MSiO$_{5/2}$-SiO$_{5/2}^-$ cluster shown in d) which derives from cluster c).}
\end{figure}
Our SIHM theory of ion-conduction builds on an approach initially devised for the conductivity of alkali borate glasses \cite{34,36}  and identifies hopping sites for conducting cations from a structural model, which in our case was used to describe application of SICA to (1-x)SiO$_2$-xM$_2$O glasses. Both (present work and ref \cite{34}) models incorporate the idea first invoked by Anderson and Stuart  \cite{37} that  activation energy for ionic conduction (E$_A$) is made of two terms-  an electrostatic energy (E$_c$) to create an ion and which determines the free carrier concentration n$_L$ , and a  network stress energy (E$_s$) that facilitates migration of ions and determines carrier mobility $\mu$. We, therefore, concentrate our efforts on estimating the energies, E$_c$ and E$_s$, as they can be directly related to the statistics of clusters and the enumeration of constraints via statistical mechanics averages. 
\par
We begin by modelling the parent solid electrolyte,(1-x)SiO$_2$-xM$_2$O , where M represents an alkali ion, and evaluate the free carrier concentration n$_L$. We will have no M$^+$ ions ( n = 0) available if we consider a pair of  Si tetrahedra, each having one non-bridging oxygen near neighbour attached to M$^+$ ions, i.e., a pair of  MSiO$_{5/2}$ tetrahedra (Fl-Fl), since there are no vacancies. However, we will have n = 1 and 2 vacancies in the case of local structure pairings of the type, MSiO$_{5/2}$-SiO$_{5/2}^-$ , and SiO$_{5/2}$-SiO$_{5/2}^{2-}$ , contributing respectively 1 and 2 M$^+$ carriers for ion transport (Figure 7). And  by knowing the probability for each of these parings p$_{jk}$, one can calculate a mean Coulombic energy over all possible pairings between Fl and St local units to obtain the carrier concentration, 
\begin{eqnarray}
n_L^{(2)}=2x\exp[-<E_c>/k_BT]
\end{eqnarray}
with the mean Coulomb energy, $<E_c>$ defined as follows,
\begin{eqnarray}
<E_c>={\frac {1}{\cal Z}}\sum_{j,k}\sum_n nE_cp_{jk}\exp[-nE_c/k_BT]
\end{eqnarray}
where E$_c$ is the Coulombic energy to extract a cation M$^+$ from an anionic site, and acts as a free parameter in the theory. ${\cal Z}$ normalises the free carrier concentration by requiring n$_L^{(2)}$ = 0 at T=0,  and n$_L^{(2)}$=2x at infinite temperature, i.e. the nominal carrier concentration. As the probability p$_{jk}$ of pairings depends on the nature of the elastic phase, the free carrier concentration n$_L^{(2)}$ can be expected to change as we go across the three elastic phases (flexible, intermediate, stressed). 
\par
Once a carrier is available, it will hop between vacant sites, and the general form for hopping rates is given by
\begin{eqnarray}
J_{ij}= \omega_{ij}\exp[-E_s/k_BT]
\end{eqnarray}           
where $\omega_{ij}$ is the attempt frequency, and we  assume it to be constant as only one type of local  species (Fl, e.g. MSiO$_{5/2}$) is involved in ionic conduction here. In fact, given the compositional region where the theory applies and where rigidity transitions are usually found \cite{17}, one does not expect to observe other Fl local species containing more than one cation (e.g. M$_2$SiO$_3$ corresponding to the metasilicate composition). In general, the attempt frequency will depend on the local environment of the hopping cations \cite{38}. The strain or migration energy E$_s$ for a hop is roughly the energy required to locally deform the network between a cation site and an available vacant site. It should therefore depend on the floppy mode energy. We assume it to not depend on the process (i,j), which is equivalent to saying that we neglect the Coulomb repulsion from vacant sites. Note that in the stressed- rigid phase, there are only few hopping events possible since the network is made up largely of stressed-rigid St-St and isostatically rigid St-Fl pairs. 
\begin{figure}
\begin{center}
\epsfig{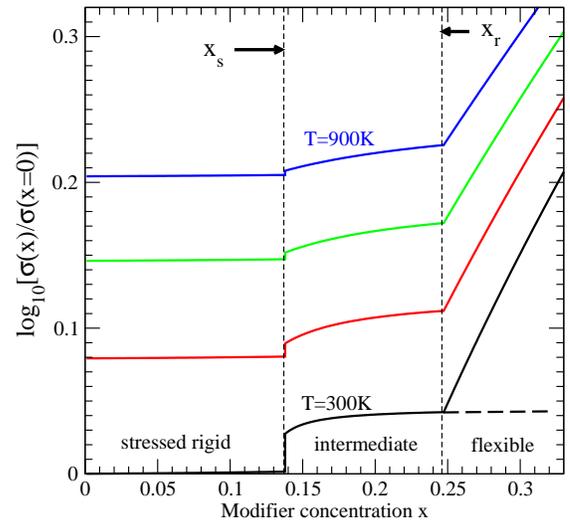}
\end{center}
\caption{(Color online) Predictions of the SIHM model for variations in ionic conductivity $\sigma$(x) for the parent solid electrolyte, (1-x)SiO$_2$-xM$_2$O, where M = Alkali ion,  using E$_c$ = 0.1 eV, and  at 4 temperatures, from bottom to top: T=300 K, 500 K, 700 K and 900 K. We have used a medium range order (ring) fraction $\eta$=0.6 which defines the IP (see text for details). Curves have been shifted along the ordinate for presentation. The broken line corresponds to the conductivity when the floppy mode contribution is neglected (see text for details).}
\end{figure}
\par
In the flexible phase (at x $>$ x$_r$ ) floppy modes proliferate as the count f$^{(2)}$ (equation (12)) becomes non-zero. In this phase, one can expect local deformations of the network to be facile, thus increasing hopping rates, and reduce the energy required to create doorways between two vacant sites. We write the strain energy as: E$_s^{flex}$=E$_s^{stress}$-$\Delta$f$^{(2)}$ where $\Delta$f$^{(2)}$ term is a typical floppy mode energy given by experiments \cite{39}. In the flexible phase, E$_s$ is reduced in relation to the stressed-rigid phase by the quantity $\Delta$f$^{(2)}$. One is then able to write a conductivity of the form:
\begin{eqnarray}
\sigma=n_L^{(2)}\mu^{(2)}=n_L^{(2)}\exp[-E_m/k_BT]
\end{eqnarray}
Results of the self-organized ion-hopping model (SIHM)  for the parent electrolyte system described above are summarized in Figure 8. In this figure we plot the variation of conductivity, $\sigma$(x) given by equ. (16), as a function of glass composition (x) at several temperatures T but at a fixed Coulomb energy E$_c$. This model calculation shows that the elastic nature (stressed rigid, intermediate, flexible) of networks alters ion-transport in profound ways. Three prominent features become apparent; (i) a minuscule conductivity change is predicted in the stressed-rigid phase, (ii) a step-like increase in $\sigma$(x) is anticipated at the stress phase boundary x = x$_s$ , and finally (iii) a logarithmic variation of $\sigma$(x) is predicted  in the flexible phase( x $>$ x$_r$). These features of our SIHM model are in rather striking accord with the observed variation of conductivity in the present AgPO$_3$-AgI  glasses (Figure 3). 
\par
The SIHM model predictions provide crucial insights into origin of ion-transport in the three elastic regimes. In the stressed-rigid phase, strain energy is high and leads to a low number of hopping possibilities, since hops are allowed only between select number of Fl-St pairs. These restrictions on hopping combined with a low free carrier concentration results in a weak dependence of $\sigma$(x)  on modifier concentration x. The order of magnitude of $\sigma$  normalized to $\sigma$(x=0), is largely determined by the magnitude of E$_c$, i.e. the interaction energy between the cation M and its corresponding anionic site.
\par
In the IP, new cation pairs (Fl-Fl, and thus hopping possibilities) are populated, and lead to a mild increase of $\sigma$ . The step-like increase in conductivity at the stress-phase boundary , x = x$_s$ (Fig 7) is a feature that is nicely observed in our experiments (Fig 3a), and provides an internal consistency to the description advanced here. Theory tells us that the jump in  $\sigma$ at stress transition depends on T or inverse E$_c$ at fixed T, (see equ. (14)). The location of the stress transition is a network property and it depends on the IP, i.e., aspects of medium range structure. As in other quantities observed \cite{8}, the first order jump in $\sigma$(x) near x = x$_s$ is a manifestation of deep structural changes in a glass network which contributes to a cusp in the configurational entropy \cite{33}. And as a final comment on the subject, we recognize that at a fixed temperature, large values for the Coulomb energy E$_c$ will lead, in general, to large jumps in conductivity at the stress transition. Here theory predicts that jumps in conductivity near the first order stress transition (x = x$_s$) will be more conspicuous with heavier cations (such as Ag or K) than with the lighter ones (Li). 
\par
Glasses become flexible at x $>$ x$_r$ as floppy modes start to proliferate and decrease the strain energy barrier. Electrical conductivity displays a second threshold at the rigidity transition x=x$_r$ , as the slope d$\sigma$/dx changes abruptly at the transition, and the variation of $\sigma$(x) becomes exponential above the threshold x$_r$ . Network flexibility clearly promotes ionic conductivity. In the flexible phase an increase in the available degrees of freedom appears to facilitate local network deformations, thus creating pathways for conduction. Furthermore, the present results also suggest that the increase in the number of possible hopping processes in the IP (region between x$_s$ and x$_r$ in Fig.7) is small and tends to produce saturation upon increasing modifier content (broken line in Fig. 7). This is to be compared to the dramatic increase of $\sigma$  that mostly arises in the flexible phase as floppy modes proliferate, f$^{(2)}\neq$ 0 , and  E$_s$ decreases sharply (Fig 9).

\subsection{Coulomb energy (E$_c$) and network stress energy (E$_s$) in AgPO$_3$-AgI}
 
\subsubsection{Coulomb energy E$_c$ and Ag$^+$ ion mobilities}

The Coulomb energy E$_c$ represents the energy to create a Ag$^+$ ion that would contribute to ionic conduction in the AgPO$_3$-AgI electrolytes glass of interest. In its simplest form \cite{14}, \cite{40} it can be written as  
\begin{eqnarray}
E_c={\frac {e^2}{4\pi\varepsilon_\infty\varepsilon_0}}{\frac {1}{R_{Ag-X}}}
\end{eqnarray}
where R$_{Ag-X}$ and $\varepsilon_\infty$ represent respectively an average Ag-X ( X = I, O) nearest-neighbour bond length and the high frequency permittivity. From the derived frequency dependence of the permittivity using the electric Modulus formalism, earlier in section IV A, we deduced the variation of $\varepsilon_\infty$(x) in the inset of Fig. 5. The quantity relates to the dielectric response in the immediate vicinity of a silver cation, and permits one to estimate the Coulombic barrier needed to create a mobile Ag$^+$ cation. In the estimate of E$_c$ we use an  average of Ag-I, and Ag-O bond length of 2.43 $\AA$ and 2.85 $\AA$ for R$_{Ag-X}$ in equation (16), and note that the R$_{Ag-I}$ bond length remains largely constant with glass composition \cite{41}. Furthermore, corrections due to many- body- effects arising from Ag$^+$-Ag$^+$ interactions can also be taken into account in equation (16) as suggested elsewhere \cite{14}. The variations of E$_c$(x) thus largely results from that of $\varepsilon_\infty$(x) (Figure 5 inset), and these are reflected in the plot of Figure 9. We find that E$_c$ varies between 0.05 eV to 0.10 eV across the broad range of compositions in the present glasses.  
\begin{figure}
\begin{center}
\epsfig{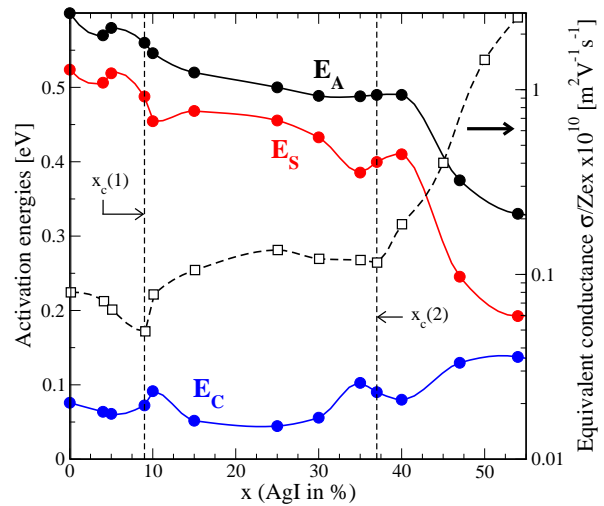}
\end{center}
\caption{Variations in the activation energy E$_A$ (same as figure 2), Coulomb energy E$_c$ , strain energy E$_s$ = E$_A$-E$_c$  and equivalent conductance at 245 K determined  from the conductivity data of Fig. 3 for (AgPO$_3$)$_{1-x}$-(AgI)$_x$  glasses.}
\end{figure}
\par 
How do the present estimates of E$_c$ compare to earlier ones in other systems ? Other groups report similar findings; in (Na$_2$O)$_x$(SiO$_2$)$_{1-x}$ glasses, Greaves and Ngai \cite{14} estimate E$_c$ near 0.1eV to 0.2 eV, and the activation energies E$_A$ ($\simeq$ 0.6 eV). In (K$_2$O)$_x$(GeO$_2$)$_{1-x}$ glasses, Jain et al. \cite{40} estimate E$_c$ to be $\simeq$ 0.5 eV, typically 30\% of E$_A$ ($\simeq$ 1.5 eV) at x $<$ 0.10. Reverse Monte Carlo Models of (AgPO$_3$)$_x$(AgI)$_{1-x}$ glasses by Wicks et al. have also emphasized E$_c$ to be small \cite{42}, and  E$_s$ to be the  major contributor to E$_A$. A similar conclusion is reached by Adams and Swenson \cite{43}, who explicitly reveal that strain energy dominates ionic conductivity. The present findings on E$_c$ highlight once again that the low value of E$_c$ appears to be generic feature of solid electrolytes. 
\par
$^{109}$Ag NMR studies \cite{27} of the base AgPO$_3$ glass show the  NMR resonance to be rather broad, fully consistent with its low conductivity ($\sigma$ $\simeq$ 10$^{-9}~\Omega^{-1}$.cm$^{-1}$, see Fig 3), in which the Ag$^+$  cations serve as compensating centers. Furthermore, the low value for  E$_c$ nearly 10\% of the total activation energy E$_A$ found in the present AgPO$_3$-AgI glassy alloys (Fig.9) suggests that the number of free Ag$^+$ carriers varies linearly with the macroscopic AgI content 'x' of the alloys.  For this reason, the equivalent conductance $\sigma$/Zex, i.e. the change in conductivity per mole \% of AgI, serves a good measure of ionic mobility (equation 1). Results on variation of the equivalent conductance are plotted in Fig. 10, right axis. Here one can see that while the nature of the elastic phases weakly affects E$_c$, and thus the free carrier density, they induce profound variations in the equivalent conductance, and thus carrier mobility. In particular, the two thresholds in equivalent conductance coincide with those noted earlier in the variations of $\Delta H_{nr}$(x) and $\sigma$(x) (Fig.3), which serve as benchmarks of the three elastic phases of the present glasses. In the flexible phase ( x $>$ 38\%) , we observe a striking increase by a factor of nearly 20! in the 40\% $<$ x $<$ 53\% range.
\par
$^{109}$Ag NMR studies \cite{27} have shown that the number of mobile carriers is nearly equal to the number of silver ions introduced by AgI. Using the simple relationship between  $\sigma$  and mobility (Equation (1)) Mustarelli et al. \cite{27} estimated the variation in mobility $\mu$(x) of carriers. Their data are also reproduced in Fig. 10 as the red filled circles. The corresponding equivalent conductance data on our samples are plotted in Fig. 10 as the open square data points. It is clear that the mobility in the present samples are at least an order of magnitude lower than in the samples of Mustarelli et al. \cite{27}.  
\par
We believe these differences in $\mu$(x) between the two sets of samples derive from sample makeup.  The glass transition temperatures T$_g$(x) are a sensitive measure of bonded water in the base glass, AgPO$_3$, as described earlier \cite{11}. Presence of bonded water in the base glass depolymerizes the P-O-P chain network and softens the network. The samples of Mustarelli et al.\cite{27} show their base glass (AgPO$_3$) T$_g$ = 189$^o$C , about 60$^o$C lower than the T$_g$ of our driest base glass of 254$^o$C. Alloying AgI in the base glass will soften it further. Furthermore, in our samples we do not observe evidence of $\beta$-AgI segregation in the 50\% $<$ x $<$ 80\% composition range as noted by Mustarelli et al. \cite{27} in their samples. The order of magnitude larger mobilities observed by Mustarelli et al., most likely, derive from the flexible nature of glass compositions promoted by the presence of additional bonded water. 
\par
\begin{figure}
\begin{center}
\epsfig{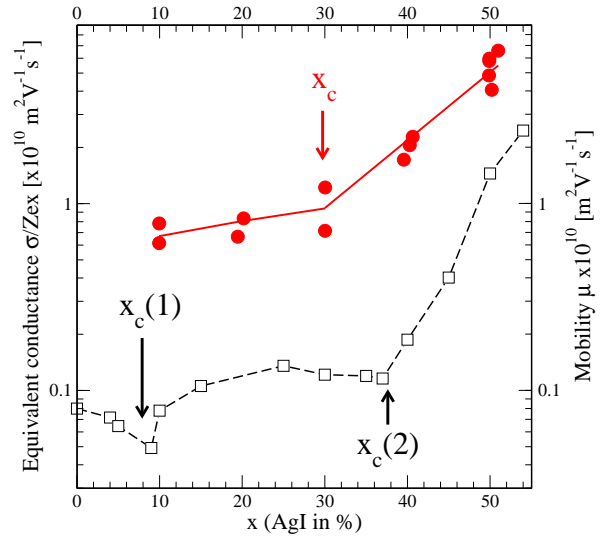}
\end{center}
\caption{(Color online) Equivalent conductance $\sigma$/Zex (open squares, broken line, same as Fig. 7) and silver mobility $\mu$ (red circles, right axis) determined from a free carrier rate NMR estimation (Mustarelli et al. Ref. \cite{27}).}
\end{figure}
In our samples, we found \cite{8} the IP composition range to be rather sensitive to water content of samples.  For example, the  IP-width nearly halved (28\% to 15\%) and the IP centroid moved up in AgI concentration from 24\% to 28\%  in going from samples of set A (driest) to those of set B (drier).  In samples of set B, the T$_g$ of the base glass was 220$^o$C, about 34$^o$C lower than in samples of set A (Tg = 254$^o$C). And possibly, the solitary conductivity (or mobility) threshold observed near x = 30\% by Mustarelli et al.\cite{27} (Fig. 10)  and separately by Mangion and Johari \cite{28} in their samples may well represent the extremal case of a  completely collapsed IP at its centroid near 30\% in their samples giving rise to only one elastic threshold with the stress- and rigidity- transitions merging.  These data highlight the importance of sample synthesis to establish the intrinsic electrical and elastic behavior of these glasses. The role of water traces in collapsing the IP in the present solid electrolyte glass is reminiscent of a collapse \cite{47} of the IP  observed in Raman scattering of a traditional covalent Ge$_x$Se$_{1-x}$  glass, measured as a function of laser power-density used to excite the scattering. In both instances, the self-organized state with a characteristic intermediate range order formed, is irrevocably changed to a random network structure, in one case by OH dangling ends replacing bridging O sites and splicing the network, and in the other case by rapid switching of lone pair bearing Se centered covalent bonds by the action of near band gap light serving as an optical pump.     

\subsubsection{Network Stress energy E$_s$}

Estimates in E$_c$(x) permit obtaining variations in the network stress energy E$_s$ (=E$_A$-E$_c$) from the measured activation energies E$_A$(x) (Figs. 1 and 2). Variations in the activation energies E$_A$(x), and E$_s$(x) are plotted in Figure 8 for the reader's convenience.  A perusal of these data clearly reveals that the drastic increase of electrical conductivity in the 40\% $<$ x $<$ 60\% range is tied to a two-fold reduction in E$_s$(x) (Fig. 9). Thus, even though the concentration of the electrolyte salt additive (AgI) increases linearly with x, changes in conductivity are not only non-linear in x, but display discontinuities at the stress (x$_c$(1) = 9\%) and the rigidity (x$_c$(2)= 38\%) elastic phase boundaries as discussed earlier.
\par
The central finding to emerge from an analysis of our electrical conductivity results is that  fast-ion conduction in dry AgPO$_3$-AgI glasses is largely controlled by the elastic nature of their backbones. Recent work on alkali-earth silicates also shows evidence of conductivities increasing precipitously once glasses become elastically flexible at a threshold additive concentration \cite{29,45}. 

\subsection{Kohlrausch stretched exponent and network dimensionality in AgPO$_3$-AgI glasses}

The Kohlrausch exponent usually reflects the degree of cooperativity of mobile ions in the glass, the smaller the $\beta$ the larger the collective behaviour of  cation motion. From what is seen in Fig. 6, one can state that collective behaviour is enhanced in the intermediate phase.
\par
We note that variation of the Kohlrausch exponent $\beta$(x) displays a trend that is similar to that of the non-reversing heat flow (Fig.3b) , with values of $\beta$ of about 0.66 in the stressed-rigid and flexible phases, and a value of about 0.52 in the IP. These $\beta$(x) data contrast with previous findings of the exponent in the Ge$_x$As$_x$Se$_{1-2x}$ ternary glasses measured in flexure measurements where one found \cite{46} that $\beta$ steadily increases in going from the flexible to the stressed-rigid phases. In the latter it converged to a value of 0.60 for a network mean coordination number larger than 2.4 corresponding to intermediate and stressed-rigid phase glasses. 
\par
The stretched exponent $\beta$ approaches 1 at high temperatures (T $>$ T$_g$) , and reduces as T decreases to T$_g$, i.e., $\beta$(T$\rightarrow$ T$_g$) $<$ 1. The reduction of $\beta$ usually reflects contraction of configuration space in the supercooled liquid and its eventual stabilization \cite{48} due to structural arrest near T=T$_g$. On this basis at a fixed network dimensionality, one should expect that an increase in the number of Lagrangian constraints (or decrease of AgI content) will contract the configuration space, leading to lower values for $\beta$. Fig. 6 obviously does not follow this anticipated behaviour as the dimensionality is changing with modifier content. Models of traps \cite{48}-\cite{51}  for tracer diffusion in a d-dimensional lattice show that $\beta$ is related to the dimensionality d via:
\begin{eqnarray}
\beta={\frac {d}{d+2}}
\end{eqnarray}
which is satisfied for various molecular supercooled liquids and glasses \cite{51} having either a 3D ($\beta$ = 3/5) or a 2D relaxation ($\beta$ = 1/2). However, in the case where an internal structural dynamics takes place, the dimensionality d of networks must be replaced by an effective one, d$_{eff}$, involved in relaxation. This happens when Coulomb forces are present as the motion of the carriers is hindered by the tendency towards local charge neutrality, determining an effective dimensionality d$_{eff}$ of the configuration space in which relaxation takes place \cite{51}. In this case, the effective dimensionality becomes, 
\begin{eqnarray}
d=d_{eff}{\frac {N_d}{N_d+N_c}}
\end{eqnarray}
where N$_d$ is the number of degrees of freedom in laboratory space and N$_c$ is the number of mechanical constraints, estimated from short-range interactions. In the IP where $N_c\simeq 3$, $\beta$ = 1/2 (Fig.6), equation (17) implies d = 2, resulting into a d$_{eff}$ = 1 from equation (18). Outside the IP, $\beta$ = 3/5 (Fig.6), equation (17) implies d = 3, resulting into a  d$_{eff}$ = 1.5 from equation (18).  Thus, presence of Coulomb interactions in AgPO$_3$-AgI glasses, suggests filamentary (d$_{eff}$ = 1) diffusion of Ag$^+$ carriers in the IP, giving rise to percolative pathways in the self-organized structure. This conclusion is supported by numerical simulations \cite{42}, which show that isolated sub-diffusive regions exist at low AgI concentration and these percolate as filaments at increased AgI concentrations in the 20\% $<$ x $<$ 30\% range. 

\section{Concluding remarks}
 
Dry and slow cooled (1-x)AgPO$_3$-xAgI glasses  synthesized over a broad range of compositions,  0 $<$ x $<$ 53\% , are investigated in ac conductivity measurements as a function of  composition,  temperature and frequency. From these data compositional trends in activation energy for conductivity E$_A$(x), Coulomb energy E$_c$(x) for Ag$^+$ ion creation, Kohlrausch stretched exponent $\beta$(x), low frequency permittivity $\varepsilon_s$(x) and high-frequency permitivity $\varepsilon_\infty$(x) are deduced. By combining rigidity theory with SICA and statistical mechanics, we have derived the analytic properties of the conductivity response function in a Ag based solid electrolyte. The topological model  reproduces  observed trends in the compositional variation of $\sigma$(x), including the observation of a step-like jump as glasses enter the Intermediate phase near  x$_c$(1)=9\%,  and an exponential  increase when glasses become flexible  at x $>$ x$_c$(2)=38\%. Since E$_c$ is found to be small compared to network strain energy (E$_s$), we conclude that free carrier concentrations  increase linearly with  nominal AgI concentrations, while the super-linear variation of  fast-ion conduction can be traced to changes in carrier mobility induced by an elastic softening of network structure.  By better understanding the role played by the "alloyed network" on fast-ion conduction in a specific Ag based solid electrolyte glass, we have also taken a step towards better understanding ion-transport in several other types of  solid electrolyte systems,  including  modified oxides and  modified  chalcogenides, systems that are also  expected to display a parallel  conductivity response as a function of network connectivity.  Variations in $\beta$(x) reveal a square- well like variation with walls localized near x$_c$(1) and  x$_c$(2), and suggest filamentary ( quasi 1D) conduction in the Intermediate Phase ( x$_c$(1) $<$ x $<$  x$_c$(2)) but a dimensionality larger than 1 outside the IP. The present findings, characteristic of dry samples and showing two elastic thresholds, may possibly go over to only one elastic threshold in wet samples in which the IP completely collapses as suggested by earlier work in the field.

\par
{\bf Acknowledgements}
\par
It is a pleasure to acknowledge discussions with Patrick Simon and Ping Chen during the course of this work.  This work is supported in part by NSF Grant DMR 04-56472 and DMR 08-53957.

\end{document}